\documentclass[hyper]{prop2015}
\usepackage[english]{babel}
\usepackage{amsmath,bbm}

\def\beq{\begin{equation}}
\def\eeq{\end{equation}}

\def    \bea    {\begin{eqnarray}}
\def    \eea    {\end{eqnarray}}

\newcommand{\Gfour}{{SL}(5)}
\newcommand{\Gfive}{{SO}(5,5)}
\newcommand{\Gsix}{E_{6(6)}}
\newcommand{\Gseven}{E_{7(7)}}
\newcommand{\Geight}{E_{8(8)}}

\newcommand{\Hfour}{{SO}(5)}
\newcommand{\Hfive}{{SO}(5)\times {SO}(5)}
\newcommand{\Hsix}{{USp}(8)}
\newcommand{\Hseven}{{SU}(8)}
\newcommand{\Height}{{SO}(16)}

\newcommand{\HfourLor}{{SO}(2,3)}
\newcommand{\HfiveLor}{{SO}(5, \mathbbm{C})}
\newcommand{\HsixLor}{{USp}(4,4)}
\newcommand{\HsevenLor}{{SU}^*(8)}
\newcommand{\HeightLor}{{SO}^*(16)}
\category{Proceedings}


\subtitle{\href{http://www.maths.dur.ac.uk/lms/109/index.html}{LMS/EPSRC Durham Symposium on Higher Structures in M-Theory}}
\title{A Kaluza--Klein Approach to Double and Exceptional Field Theory}
\author[D.~S.~Berman]{David S. Berman\inst{a,}\footnote{Corresponding author e-mail:~\href{mailto:D.S.Berman@qmul.ac.uk}{\textsf{D.S.Berman@qmul.ac.uk}}}}
\address[1]{School of Physics and Astronomy, Queen Mary University of London, 327 Mile End Road, London E1 4NS, United Kingdom}

\shortauthors{D. S. Berman}
\begin{abstract}
We examine the challenge of viewing all the fields in supergravity as arising from a Kaluza--Klein like dimensional reduction of some higher-dimensional theory. This gives rise to what is known as exceptional field theory or double field theory. A particular emphasis is placed on following the  Kaluza--Klein intuition leading to the identification of charged states and a reinterpretation of the central charges. We further give a description of the novel extended geometry as a generalised phase space and the relationship to string and M-theory theory and the notion of quantization. 
\end{abstract}
\shortabstract
\begin{document}
\maketitle



\section{Introduction}

\subsection{Some small history}

This paper as a whole follows a very non-historically accurate approach to double and exceptional field theory based on a Kaluza--Klein approach to supergravity. Before beginning this chain of logic let us first briefly describe some of the history of the subject. Almost 30 years ago Michael Duff \cite{Duff:1989tf} developed a string world-sheet theory where T-duality appeared as a manifest symmetry. Subsequently Arkady Tseytlin \cite{Tseytlin:1990nb,Tseytlin:1990va} described different aspects of a string in a doubled space-time with manifest T-duality before Warren Siegel \cite{Siegel:1993bj} described a sophisticated world-sheet theory with manifest $O(d,d)$ symmetry. After a break of some years, in 2009, Hull and Zwiebach \cite{Hull:2009mi,Hull:2009zb} and then with Hohm \cite{Hohm:2010jy,Hohm:2010pp} examined a truncation of closed string field theory keeping only the momentum and winding modes of the string field and produced a string background with twice the number of coordinates. This theory is known as double field theory (DFT). Somewhat in parallel, the Korean group \cite{Jeon:2010rw,Jeon:2011cn } developed what is now called the semi-covariant formalism.

In parallel developments, in the work of Cremmer, Julia and Scherk \cite{Cremmer:1978km}, eleven-dimensional supergravity reduced on a $d$-dimensional torus was shown to exhibit an $E_d$ exceptional group of global symmetries. Later with the advent of M-theory \cite{Hull:1994ys,Witten:1995ex} this $E_d$-symmetry was extended to a duality of string theories known as U-duality that combined T-duality with S-duality. Various works then attempted to reformulate supergravity theories such that exceptional symmetry would become a manifest symmetry of the theory, notably in the numerous works of West \cite{West:2001as}, Nicolai \cite{Gebert:1994zq} and others. In \cite{Hillmann:2009zf}, by extending the number of dimensions, the group $E_7$ was made into a manifest symmetry and in \cite{Berman:2010is} the group $SL(5)$ was similarly made manifest again by extending the number of dimensions. There have been other numerous developments, perhaps most notably the role of generalised geometry in this scheme has been developed in \cite{Coimbra:2011ky,Coimbra:2012af}. The main thrust of this paper is to focus on the extra coordinates and so this will not be our approach. 

Crucially, as we can see from the above narrative, the duality symmetries in string and M-theory were at the heart of the developments for these theories. We wish to emphasize that in the approach described here this is only a by-product of these theories and one should not see DFT or exceptional field theory (EFT) as a theory to make duality symmetries manifest. To do so would make these theories redundant for backgrounds without isometries (T-duality in its usual form requires isometries of the background) and this is not true. 

\subsection{Kaluza--Klein theory}
Let us begin with a review of traditional Kaluza--Klein theory. This is so we can give a prescription for a series of steps that we will emulate later for all the bosonic  fields in supergravity and so produce double or exceptional field theory.
The starting point is Einstein--Maxwell theory in four dimensions coupled to a scalar field. The field content is thus the metric $g_{\mu \nu}$, the one-form vector potential $A_\mu$ and the scalar field $\phi$. The action for these fields is given by
\begin{equation}
S= \int {\rm d}^4 x \sqrt{-g}\,e^\phi\left( R(g) - \tfrac{1}{4} F_{\mu \nu} F^{\mu \nu} - \tfrac{1}{2} \partial_\mu \phi \partial^\mu \phi\right)~, \label{4dem}
\end{equation}
where $F_{\mu \nu}= \partial_\mu A_\nu -\partial_\nu A_\mu$ is the field strength for $A_\mu$. The local symmetries of the theory are given by diffeomorphisms, as described by the Lie derivative,
\begin{equation}
L_V U ^\mu = V^\rho \partial_\rho U^\mu + \partial_\rho V^\mu  U^\rho\,,
\end{equation}
and the gauge transformations of $A_\mu$,
\begin{equation}
\delta A_\mu = \partial_\mu \chi \, .
\end{equation}
Thus there are five parameters for the local symmetries: the four vector that generates the diffeomorphisms, $V^\mu$, and the single scalar $\chi$ that generates the gauge transformations of $A_\mu$. 

The Kaluza--Klein idea may then be expressed as follows: given there are five parameters for the local symmetries, can one combine them to form a five vector $\hat{V}^{\hat{\mu}}$ such that diffeomorphisms generated by this five vector acting on some five-dimensional metric, $\hat{g}_{\hat{\mu }\hat{\nu}}$ will reproduce the four-dimensional local transformations described above. (We take hatted objects to be five-dimensional and $\hat{\mu}=(\mu,5)$.)  

The challenge then is to find the five-dimensional metric that meets this criteria, and the answer can be found to be:
\begin{gather}
\hat{g}_{\hat{\mu }\hat{\nu}}=\begin{bmatrix} g_{\mu \nu} + \phi^2 A_\mu A_\nu & \phi^2 A_\nu \\ \phi^2 A_\mu & \phi^2 \end{bmatrix} \, . \label{kkmetric}
\end{gather}
For the five-dimensional theory to have five-dimensional diffeomorphism invariance implies that it should be described by the five-dimensional Einstein--Hilbert action,
\begin{align}
S_5= \int {\rm d}^5x \sqrt{-\hat{g}}\, R(\hat{g}) \, .  \label{Kaluza--Kleinaction}
\end{align}
Inserting  the ansatz (\ref{kkmetric}) into this five-dimensional action reproduces the four-dimensional action (\ref{4dem}) provided that the fields are independent of the new fifth dimension i.e. for all fields and gauged transformations:
\begin{align}
\partial_{x^5}  {} \, \,= 0  \, .  \label{Kaluza--Kleinconstraint}
\end{align}
Another way to think of this is that the original local transformations only depended on four dimensions and so to reproduce this we need the constraint (\ref{Kaluza--Kleinconstraint}) to apply to the parameters generating the local  transformations.

Thus, the five-dimensional diffeomorphisms restric\-ted to four dimensions by (\ref{Kaluza--Kleinconstraint}) are equivalent to four-dimen\-sional diffeomorphisms and one-form gauge transformations. Strictly speaking we have only seen this infinitesimal transformations and have not examined so called `large' transformations i.e.~diffeomorphisms or gauge transformations that are finite and not connected to the identity. 

In some sense we were lucky in that the combination of diffeomorphisms and one-form transformations nicely combine into five-dimensional diffeomorphisms (it is hard to imagine an a priori argument that this had to be the case without already knowing about Kaluza--Klein theory). In the sections that follow we will have to be a little more creative since to combine diffeomorphism with $p$-form gauge transformations will not produce usual diffeomorphisms in a higher-dimensional theory. Before doing this, let us examine the Kaluza--Klein approach some more as  this will provide a guiding hand later.

Now that we have the higher-dimensional theory  (\ref{Kaluza--Kleinaction}), the metric ansatz (\ref{kkmetric}) and the constraint for the reduction (\ref{Kaluza--Kleinconstraint}) the next step is to look for the origin of the electrically charged states in the theory. So far the action (\ref{4dem}) describes Einstein--Maxwell theory with no currents. How can we add electric sources from the new higher-dimensional perspective?

The answer is given quickly by calculating the geodesic equations for a probe particle in the five-dimensional theory with the ansatz (\ref{kkmetric}). Doing this, one discovers that the Lorentz force law for electric charges is recovered provided one identifies the four-dimensional electric charge with the derivative in the fifth dimension. One then uses the usual relationship between the probe particle wave function and the momentum operator as a derivative to write:
\begin{align}
i\partial_5= P_5 = Q_e \, .
\end{align}
Thus objects with momentum in the fifth direction will appear as electric charges from the four-dimensional perspective. A light-like object in five dimensions will obey:
\begin{align}
P^{\hat{\mu}} P_{\hat{\mu}} =0  \,   \qquad P^5 P_5 + P^{\mu} P_{\mu} =0 \, .
\end{align}
Then, since the four-dimensional on-shell relation is $ P^\mu P_\mu=-M^2$, this implies that $P_5=M$ and thus the BPS condition:
\begin{align}
 M=Q_e  \, .
\end{align}
At this point one may ask about electric charge quantisation. If one requires $Q_e$ to be quantised then this implies that the extra Kaluza--Klein direction is compact, typically one takes this direction to be a circle of radius $R$. Then the momentum operator $P_5$ will have a discrete spectrum with $P_5 = {\hbar} \frac{n}{R}$ where $n$ is an integer. This is the usual way Kaluza--Klein quantises electric charge, which differs slightly from the usual Dirac quantisation in that there is no mentioning of magnetic charges.
The alert reader may feel slightly dissatisfied with the fact that although the electromagnetic field is geometric the charges are not. This may be remedied as follows. The clue is to take seriously the identification of momentum with charge and mass. Now to realise charge and mass using only a gravitational degrees of freedom one then seeks a gravitational solution with momentum in the Kaluza--Klein direction. Momentum is then thought of in the ADM \cite{Arnowitt:1960es} sense. To write a solution whose ADM momentum is $P_5$ one needs a Killing symmetry in that direction. To require the mass to be also given by $P_5$ then also needs a Killing direction in time and the solution be a null wave such that $P_5=P_0$. Such solutions are known and are called pp-waves. The solution is as follows:
\begin{align}
{\rm d}s^2=-H^{-1} {\rm d}t^2 + H\big({\rm d}x^5-(H^{-1}-1){\rm d}t\big)^2 + {\rm d}y_{d-2}^2~.\label{wave}
\end{align}
The function $H$ is a harmonic function of the transverse space, $H(r)=1+ p/r$, and the parameter $p$ in the harmonic function will be the momentum.
Note that since this has a Killing symmetry along $x^5$, it is independent of the $x^5$coordinate and obeys the Kaluza--Klein constraint on the metric (\ref{Kaluza--Kleinconstraint}). This is in contrast to the usual quantum mechanical intuition whereby fields in momentum eigenstates depend exponentially on the associated coordinate. At this point, without quantum mechanics charge quantisation is mysterious from the purely gravitational point of view.
These null waves are solutions that construct for us electrically charged objects from pure gravity in five dimensions. The next natural question to ask is how one may construct magnetic charges? Following Dirac's construction of magnetic monopoles in electromagnetism indicates that instead of thinking of the Kaluza--Klein total space as a product space, $M^4 \times S^1$, one declares that the circle is fibred over the $M^4$ such that it is only $M^4 \times S^1$ locally. In fact, to construct a monopole one uses the Hopf fibration where $S^3$ is a circle fibred over an $S^2$ base,

For the monopole solution in gravity one then takes the base $S^2$ for the Hopf fibration as coming from the angular polar coordinates on the space transverse to a point. In other words, $M^4= R_t \times R_+ \times S^2$, where $R_t$ is the time direction which plays no role and $R_+$ is the radial direction. One now fibres the Kaluza--Klein circle over the $S^2$ to make $S^3$ and allow a fibration of the resulting  $S^3$ over $R_+$. This solution was found independently by Sorkin \cite{Sorkin:1983ns} and Gross and Perry \cite{Gross:1983hb}:
\begin{align}
{\rm d}s^2=-{\rm d}t^2 + H^{-1}({\rm d}z^2 +A_i {\rm d}y^i)^2 +H {\rm d}y^2 \, .
\end{align}
The field $A_i$ which controls the twist is related to the harmonic function as follows:
\begin{align}
\partial_{[i} A_{j]}= \tfrac{1}{2} \epsilon_{ij}{}^k \partial_k H \,  \, , \quad H= 1+ \frac{g}{r} \, .
\end{align}
From the perspective of four dimensions and the gauge field $A_{\mu}$, this is an object with magnetic charge $g$. Topologically the magnetic charge is the first Chern class of the circle fibration and is thus quantised. One can then use the relationship between radius of the Hopf circle and the electric charge to produce the Dirac quantisation condition:
\begin{align}
eg=2\pi n \hbar \, .
\end{align} 
The quantisation now though is topological in origin, it requires the fibre to be a circle and its twist over the base must be integer. Thus demanding a magnetic solution requires the Kaluza--Klein isometric direction to be a circle. Remarkably even though there is no metric dependence on this direction the  topology of how it is fibred gives rise to the magnetic charge.

There are then a set of obvious extensions to this idea. One can have yet more hidden dimensions, let us denote by $d$ their number. Once one has this then there is a question about their geometry. Having the $d$ hidden dimensions be a torus just gives $d$ copies of $U(1)$ gauge theories. One can do something more interesting and recover non-Abelian gauge fields and Yang--Mills theory if our $d$ dimensional space is a group manifold. So for example take the hidden space to be $S^3$ which is the group manifold of $SU(2)$. One then carries out a so called {\it{reduction}} using twist matrices often called a Scherk--Schwarz reduction. These twist matrices are essentially given by the left-invariant Maurer--Cartan one-forms on $SU(2)$. 

\subsection{M-theory}

The type IIA string low energy effective action is IIA supergravity whose bosonic fields split into the so called NS-NS sector which is the metric, dilaton and two-form potential: $g_{\mu \nu}, \phi, B_{\mu \nu}$ with $H={\rm d}B$ the field strength of $B$ and the RR sector whose fields are one-form and three-form potentials $C_\mu, C_{\mu \nu \rho}$ with field strengths $G_{(2)}={\rm d}C_{(1)}$ and $G_{(4)}={\rm d}C_{(3)}-C_{(1)} \wedge H_{(3)}$, respectively.

The action is given by: 
\begin{equation}
\begin{aligned}
S&=\int {\rm d}^{10} x \sqrt{-g} e^{-2\phi}\left(R -\tfrac{1}{12} H^2 - \tfrac{1}{2} (\partial \phi)^2 \right)\,- \\ 
&\kern.5cm-\int\left(\tfrac{1}{4}G_{(2)}^2 -  \tfrac{1}{48} G_{(4)}+\tfrac{1}{2}G_{(4)}\wedge G_{(4)} \wedge B\right) \, .  \label{IIA}
\end{aligned}
\end{equation}
The expectation value of $e^\phi$ is the string coupling $g_s$. The essence of M-theory is now to carry out the Kaluza--Klein programme and lift the metric and one-form potential $C_\mu$ to an eleven-dimensional metric. This is just a traditional Kaluza--Klein lift with:
\begin{align}
{\rm d}s_{11}^2= g^{(11)}_{\mu \nu} {\rm d}x^\mu {\rm d}x^\nu + R_{11}^2 ( {\rm d}x^{11}- C_\mu {\rm d}x^\mu )^2
\end{align}
where we have identified the RR one-form $C_\mu$ with the Kaluza--Klein vector field. The other fields must then be related as follows. A three-form potential of eleven dimensions, $C^{(11)}_3$ comes from combining the two-form, $B^{(10)}_2$ (which is the NS-NS two-form of IIA) and the  ten dimensional IIA RR three-form $C^{(10)}_3 $. That is:
\begin{align}
C^{(11)}_{11 \mu \nu} = B^{(10)}_{\mu \nu}  \, , \qquad  C^{(11)}_{\mu \nu \rho} = C^{(10)}_{\mu \nu \rho} \, .  \label{CtoB}
\end{align}
With these identifications and after a Weyl scaling of the metric then we can identify the eleven-dimensional theory as the bosonic sector of eleven-dimensional supergravity. The necessary Weyl scaling to allow this identification relates the eleven-dimensional metric in terms of the ten dimensional metric in the IIA action as follows:
\begin{align}
g^{(11)}_{\mu \nu} = \frac{1}{R_{11}} g^{(10)}_{\mu \nu} \, .
\end{align}
When the dust settles one is left with the following eleven-dimensional action:
\begin{align}
S=\int {\rm d}^{11}x \sqrt{-g} R - \int\tfrac{1}{48} G^2 - \int\tfrac{1}{6} G\wedge G \wedge C \, ,
\end{align}
where $G={\rm d}C$ is the field strength for the eleven-dimen\-sional three-form $C$.
The key aspect of the lift is that the string coupling in IIA then is given by the $R_{11}$ as follow:
\begin{align}
g_s = \left( \frac{R_{11}}{l_p} \right)^{\frac{3}{2}} \, .
\end{align}
Thus the strong coupling limit is  the limit in which the theory recovers the full eleven-dimensional symmetry. 
Again following the Kaluza--Klein intuition one sees that the charged object associated to $C_\mu$ which is the D0 brane is the momentum in the 11th direction. (The D0 is BPS which thus implies one should take a null wave in the eleventh direction.) Finally, if one twists the eleven-dimensional circle to make a Kaluza--Klein magnetic monopole using the eleventh direction then one obtains the D6 brane.

Hence, M-theory geometrises the D0 and D6 branes to become waves and monopoles in the higher-dimen\-sional theory. Much of the rich structure of M-theory comes from how the IIA fields are restructured by the eleven-dimensional lifting.
One other thing to note is that once one has the eleven-dimensional theory one can reduce in a different way. Different reductions will give different perturbative string theories.

This eleven-dimensional lifting has obviously led to a huge number of non-trivial results but in retrospect it looks like we have only done a small fraction of the job. We have combined the metric with the one-form RR field but what about the NS two-form and the other RR fields? The approach is far from being universal.
Thus in what follows we will want to combine all the $p$-form fields with the metric and so {\it{geometrise}} all the fields and make them part of a single higher-dimensional object.

\section{Lifting NS-NS supergravity, double field theory}

Now that we have seen in Kaluza--Klein theory (and M-theory) how to combine one-form gauge fields with the metric, let us move to the next simplest case of trying to combine the NS-NS two-form gauge field, $B_{\mu \nu}$, with the metric, $g_{\mu \nu}$. This NS-NS sector is common to all the supergravities whereas the spectrum of the RR fields differ between IIA and IIB and does not exist in the Heterotic or type I theories. The action in $d$ dimensions which also includes the dilaton field $\phi$ is given by:
\begin{align}
S=\int {\rm d}^{d} x \sqrt{-g} e^{-2\phi}\left(R -\tfrac{1}{12} H^2 - \tfrac{1}{2} (\partial \phi)^2\right)   \, ,  \label{NSsugra}
\end{align}
where $H={\rm d}B$ is the field strength of $B$.

The local symmetries are now the $d$-dimensional diffeomorphisms generated by a vector field $v^\mu$ through the Lie derivative as before combined with the $U(1)$ two-form gauge transformation which is generated by a one-form gauge parameter $\chi_\mu$ as follows:
\begin{align}
\delta B_{\mu \nu} = \partial_{[\mu} \chi_{\nu]} \, . \label{Bgaugetrans}
\end{align} 
Now, following our intuition from the Kaluza--Klein case we want to combine the parameters of the local symmetries into a single {\it{generalised}} vector field $V^I=(v^\mu, \chi_{\nu})$.
Immediately one sees that this requires the {\it{generalised}} vector to be $2d$-dimensional, so that $I=1,\ldots,2d$. Also, there is an unusual property that the first $d$ components of the vector are contravariant with respect to $d$-dimensional diffeomorphisms while the second $d$ components transform covariantly. Mathematically one thinks of vector fields as sections of the tangent bundle. Now for the {\it{generalised}} vector field it is a section of the direct sum of the tangent bundle and cotangent bundle. This extension of the tangent bundle of $M$ by the cotangent bundle of $M$ is often denoted by:
\begin{align}
TM \oplus T^*M
\end{align}
and is the basis for {\it{generalised geometry}} developed by Hitchin and Gualtieri \cite{Gualtieri:2003dx}. $M$ denotes the original $d$-dimensional manifold. Following our Kaluza--Klein intuition indicates that we should not only extend the tangent space but extend the space itself to $2d$ dimensions. We then introduce coordinates on this doubled space. \footnote{To be able to do so we will assume that the space is locally isomorphic to $\mathbbm{R}^{2d}$.} We take the doubled coordinates to be $X^I=(x^\mu, \tilde{x}_\mu)$ where $\tilde{x}_\nu$ are the new novel coordinates of the doubled space. The capital Latin indices $I,J$ etc. run from $1,\ldots,2d$ and the Greek indices $\mu, \nu$ etc. run from $1,\ldots,d$.

Inspired by Kaluza and Klein we then seek a so called {\it{generalised metric}} for this $2d$ dimensional extended space that will combine the fields $g_{\mu \nu}$ and $B_{\mu \nu}$ into a single generalised geometric object. This generalised metric is given by:
\begin{align}
M_{IJ} =\left( \begin{matrix} 
g_{\mu \nu} - B_\mu{}^\alpha B_{\alpha}{}_{\nu}   &  B_\mu{}^{\beta} \\
B^{\alpha}{}_\nu & g^{\alpha \beta}
\end{matrix} \right) \, .  \label{genmetric}
\end{align}
One sees it follows the Kaluza--Klein metric ansatz for $d$ additional dimensions but with the peculiarity that the new novel dimensions have been assigned a metric $g^{\mu \nu}$. The origin of this metric and its properties will be discussed  in more detail later; for now it is sufficient to see it as a natural generalisation of the Kaluza--Klein metric with $d$ extra dimensions.
 
The next step is to construct an action functional for the generalised metric (and dilaton) that under {\it{reduction}} will be equal to the supergravity action \eqref{NSsugra}. Reduction here means that we remove functional dependence on the new $d$ coordinates. That is (for now) we will demand, that:
\begin{align}
\partial_{\tilde{x}_\nu}  =0  \label{dftKaluza--Klein}
\end{align}
when acting on the fields. This is the equivalent of the Kaluza--Klein constraint (\ref{Kaluza--Kleinconstraint}). An impetuous reader might be tempted to try the Einstein--Hilbert action in $2d$ dimensions with the generalised metric (and usual dilaton). This does not give the right answer. To see why we can use our previous Kaluza--Klein calculation to quickly do the reduction of the $2d$-dimensional Einstein--Hilbert action with generalised metric, $B_{\mu}{}^{\nu}$ from the normal Kaluza--Klein perspective are $d$  vector fields, one for each new dimension. Under this reduction, the $2d$-dimensional Einstein--Hilbert action gives the $d$-dimen\-sional Einstein--Hilbert action along with the term\linebreak $-1/4 F_{\mu \nu} {}^{\alpha} F^{\mu \nu}{}_{\alpha}$ where $F_{\mu \nu}^{\alpha}= \partial_\mu B_\nu{}{}^\alpha - \partial_\nu B_\mu{}{}^\alpha$ is the field strength for $d$ one-form fields. Under the gauge transformation (\ref{Bgaugetrans}) for $B_{\mu \nu}$ this field strength $F_{\mu \nu}$ is not gauge invariant and thus the action is manifestly not symmetric under this symmetry. This is disappointing, as it means there is there not a lift of a two-form gauge theory to Riemannian geometry in higher dimensions. Or in other words, $2d$ diffeomorphisms do not contain the gauge transformations of $d$-dimensional two-form gauge transformations. This statement can be seen directly by examining the Lie derivative in $2d$ dimensions acting on the $2d$ generalised metric subject to the constraint (\ref{dftKaluza--Klein}). The usual Lie derivative generates: $\delta B_{\mu \nu} = \partial_\mu \chi_\nu$ where there is no antisymmetrisation on indices and is thus not the gauge transformation for the $B$-field. (This is just the usual transformation of $d$ $U(1)$ one-form fields.)
One could just give up now and it is perhaps this result which explains why the Kaluza--Klein lift of theories with higher form gauge fields has only recently been developed. Instead we will persevere with the knowledge that the lift will not be usual geometry but some generalisation and it will be our goal to uncover its structure.

First let us find the action by brute force. We will take a completely general two derivative action of the generalised metric and dilaton with all $2d$ indices contracted. The coefficient for each term will then be fixed by demanding that the action reduces to the correct one after imposing (\ref{dftKaluza--Klein}). Remarkably, this can be done. (Given that there are just six different terms for the alternative contractions, and thus ignoring an overall scaling of the action this means just five coefficients are fixed to give the bosonic NS-NS supergravity action. In terms of a two derivative action for the metric and $B$-field, the supergravity action has over 10 terms, thus it is very unclear that this a priori this is possible.) Fixing these coefficients (and allowing some integration by parts which implies some additional surface terms as given in \cite{Berman:2011kg}) we find the action for the generalised metric $M_{IJ}$ that does the job is:
\begin{equation}
\begin{aligned}
S&=e^{-2d}\big(\tfrac{1}{8} M^{MN} \partial_M M^{KL} \partial_N M_{KL}\,-\\
&   \kern1.5cm- \tfrac{1}{2} M^{MN} \partial_N M^{KL} \partial_L M_{MK}\,- \\ 
&\kern1.5cm- 2 \partial_M d \partial_N M^{MN} + 2 M^{MN} \partial_M \partial_N d \big)~,
\end{aligned}
\end{equation}
where we have introduced a new rescaled dilaton field $d$ related to the usual dilaton $\phi$ by $e^{-2d}=e^{-2\phi} \sqrt{g}$.
The next step is to determine the local symmetries. The idea is that the full $2d$ transformations that are generated by a generalised vector field $V^I$ will preserve some additional structure and so the Lie derivative will be deformed. The guiding principle will be that the new {\it{generalised}} Lie derivative when reduced using (\ref{dftKaluza--Klein}) should give $d$ diffeomorphisms and two-form transformations (\ref{Bgaugetrans}). We write the generalised Lie derivative in terms of the usual Lie derivative and a deformation as follows:
\begin{align}
\hat{\cal{L}}_V U^I = {\cal{L}}_V U + Y^{IJ}{}_{KL}  U^K \partial_J V^L \, , \label{genlie}
\end{align}
where ${\cal{L}}_V U^I$ is the usual Lie derivative and $Y^{IJ}{}_{KL}$ is a globally defined invariant tensor that is to be determined based on the above requirements. For the case at hand the Y-tensor is determined to be:
\begin{subequations}
\begin{align}
Y^{IJ}{}_{KL}= \eta^{IJ}\eta_{KL}~,
\end {align}
where
\begin{align}
\eta_{IJ}= \left( \begin{matrix}
 0 & 1 \\ 
 1 & 0 
\end{matrix} \right)
\end{align}
\end{subequations}
is a globally defined $O(d,d)$ tensor. A crucial property of (\ref{genlie}) is that it leaves the $\eta_{IJ}$ invariant and thus the generalised Lie derivative actually generates local continuous $O(d,d)$ transformations.
This $O(d,d)$ structure appears throughout the theory. The generalised metric (\ref{genmetric}) is in fact a representative of a coset:
\begin{align}
\frac{O(d,d)}{O(d)\times O(d)}  \, .
\end{align}
It obeys the condition:
\begin{align}
M^{IJ}=\eta^{IK} M_{KL} \eta^{LJ}~,
\end{align}
which means that one can raise or lower indices with either the generalised metric or $\eta_{IJ}$.
Essentially this $\eta$ tensor is what allows one to split the coordinates on the space into $x^\mu, \tilde{x}_\nu$. That is it allows us to have a polarisation on the space.
The next step is to examine the consistency of all of the above. First, we must check whether the action for the generalised metric  is invariant under the generalised Lie derivative. This is a non-trivial check since the action has not been constructed using covariant objects as we would in Riemannian geometry. Basically, we have the equivalent of the Ricci scalar written in terms of partial derivatives, the metric and its inverse and so the symmetry under local transformations is far from obvious. Fortunately, the action is indeed invariant under transformations (\ref{genlie}).
Then we must examine the consistency of the local symmetry itself and study the algebra of the transformations generated by the generalised Lie derivative. First, let us recall that for the usual Lie derivative:
\begin{align}
\{{\cal{L}}_U,{\cal{L}}_V\} = {\cal{L}}_{[U,V]}  \, .
\end{align}
Now we discover that:
\begin{align}
\{{\hat{\cal{L}}}_U,{\hat{\cal{L}}}_V\} = {\hat{\cal{L}}}_{[U,V]_C}+ a \eta^{IJ}\partial_I \partial_J~, \label{alg}
\end{align}
where the hatted terms are the generalised Lie derivates given by (\ref{genlie}) and the $[.,.]_C$ is the so called `C-bracket which is defined as:
\begin{align}
[U,V]^M_C=  U^P\partial_P V^M - \tfrac{1}{2} \eta^{MN}  \eta_{PQ} U^P \partial_N V^Q - (U \leftrightarrow V   )    \, .
\end{align}
The second term in the algebra (\ref{alg}) prevents it from closing. At this point one should worry. In addition to this the above `algebra' of generalised Lie derivatives (\ref{alg}) does not obey the Jacobi identity. The resolution to these issues is that one must impose a covariant constraint on all fields given by
\begin{align}
\eta ^{IJ} \partial_I \partial_J \phi = 0 \, \label{weakc}
\end{align}
with $\phi$ any field in the theory.
When this constraint is applied then the algebra closes. As we will see later many other issues will also disappear with this constraint applied. 
Writing out the constraint in terms of the coordinates $x^\mu,\tilde{x}_\nu$,
\begin{align}
\partial_{x^\mu} \partial_{\tilde{x}_\mu} {\phi}= \partial_{\mu} \partial^{\tilde{\mu}}   \phi=0  \, .
\end{align}
This is obviously solved by the Kaluza--Klein like constraint that we have been using to construct the theory:
\begin{align}
\partial_{\tilde{x}_\mu} \phi= 0 \, . \label{strongc}
\end{align}
Here though we see the first clear departure from the Kaluza--Klein paradigm,  the reduction of the theory by definition obeys (\ref{strongc}) but for Kaluza--Klein theory the unreduced theory has no constraints. Here the `unreduced theory' still obeys a (quite stringent) constraint, (\ref{weakc}). Note that the constraint (\ref{weakc}) is weaker than the constraint (\ref{strongc}) and so it is certainly still a reduction of the theory. Working with the full theory that only obeys the weaker constraint is still a key challenge for Double Field theorists. There is much more to be said here about how if one expects to implement the constraint on products of fields then the so called weak constraint (\ref{weakc})   becomes the following `strong constraint':
\begin{align}
\eta^{IJ} \partial_I \phi  \partial_J \psi =0  \label{strongC} \, .
\end{align}
Straight away however one see that the above constraint can be solved in the larger theory with the alternative choice:
\begin{align}
\partial_{x^\mu} \phi =0  \, , \label{altsec}
\end{align}
so that now the fields will depend on $\tilde{x}_\nu$ only. There is no Kaluza--Klein equivalent statement.
How can we interpret this choice of solution of the strong constraint?
First let us discuss how we identify our physical space-time within the doubled space. We solve the strong constraint to discover that our generalised metric and dilaton depend on a set of coordinates. We then identify those coordinates as corresponding to those of our space-time. We then make what is known as a `choice of section' (the strong constraint is also called the section condition by some). The choice of section is the identification of a $d$-dimensional subspace within the doubled $2d$-dimensional space in which the double field theory is constructed. When we solve the strong constraint and determine which coordinates the fields depend on then this gives us a canonical choice for the choice of section, that is subspace, for which those are the coordinates. Now, if one uses (\ref{altsec}) as a choice for solving the strong constraint then the canonical choice of section is the space spanned by the $\tilde{x}_\nu$ coordinates. Reducing the theory to these choice of section then reproduces something unexpected. One again reduces to $d$-dimensional NS-NS supergravity. 

\subsection{T-duality in double field theory}

We see in the above that there are thus a variety of reductions of the double field theory (DFT) that give supergravity. Above we chose all $x$ coordinates or all $\tilde{x}$ coordinates but obviously one can make a mixed set provided the strong constraint is obeyed. In each case one lands on $d$-dimensional supergravity.
The independence of the generalised metric on $d$ coordinates is like having $d$ isometries (although written in non-covariant language). Now let us consider the case where we have additional isometries and the generalised metric depends on less than $d$ coordinates.  To be concrete let us consider a particular background given by a metric and $B$-field in which there was no dependence on some particular coordinate which we will label $z$ nor on its canonical pair $\tilde{z}$ (the $\eta$ tensor gives a pairing between every coordinate $x$ and every $\tilde{x}$. This background obviously solves the strong constraint and even does so degenerately for $z$ and $\tilde{z}$ in that the fields do not depend on either of them. Now the section condition prescription to determine the $d$-dimensional subspace of the $2d$ space is ambiguous. One could take the choice of section to be the space equipped with coordinates $x,z$ or the space with coordinates $x, \tilde{z}$. Since neither coordinate appears in the solution it can be included trivially in our choice of what we call space-time. However when we do this that means we will identify different components of the generalised metric with the space-time metric. Thus with a single isometry there is a $\mathbbm{Z}_2$ ambiguity. The different choices will then imply different choices of metric for space-time (and $B$-field in higher dimensions). These different choices are related through what is called T-duality and the rules determining   how the metric and $B$-field transform when there is an isometry are known as the Buscher rules. All of this is manifest in the double field theory as a consequence of the ambiguity in section choice in the presence of an isometry. For $d$ isometries, the transformations form an $O(d,d)$ group.  This is immediately realised in double field theory as a linear transformation on the generalised metric. It is in this sense that double field theory makes `duality manifest'. It is important to realise though that without isometries there is no duality but there is still the local $O(d,d)$ of generalised Lie derivatives that captures the local symmetries of the theory. The shift in the dilaton under T-duality is captured by relating the DFT dilaton to the usual dilaton and requiring the usual measure for the $d$-dimensional section being $\sqrt{g}$. Relating the two dual theories then produces the usual dilaton shift. Note that the generalised metric has unit determinant and so does not contribute to the measure; in DFT the measure is the DFT dilaton. Once we pick a section and choose our $d$-dimensional space-time then we need a measure on this space which we do with a field redefinition of the DFT dilaton as follows,
\begin{align}
e^{-2d } = \sqrt{g} e^{-2\phi}  \, .
\end{align}
 For different choice of section this will be different giving rise to the induced dilation transformation between the dilaton $\phi$ and its dual $\tilde{\phi}$,
\begin{align}
e^{-2\phi}=e^{-2\tilde{\phi}} {g} \, .
\end{align}

\subsection{Charged states}

Again we follow the Kaluza--Klein intuition and look at the charged states of the theory. Rather than examine probes of the background and their geodesic equation the approach we will take will be to look for solutions that have ADM type of charges. First, we should develop an ADM formalism for DFT (this has been done by a set of people \cite{Blair:2015eba,Park:2015bza,Naseer:2015fba}). Then as expected the ADM momentum in $p_{\tilde{x}_\nu}$ will be associated with the electric charge of the three-form field strength:
\begin{align}
Q_e= \int   {}^* H  \, .
\end{align}
The object that carries such an electric charge is the string. (The direction of the momentum in the $\tilde{x}$ space corresponds to the orientation of the string in usual space-time.)

We now seek a solution to the DFT equations of motion that has $p_{\tilde{x}_\nu}$ momentum and following the Kaluza--Klein approach it will have the structure of a null plane fronted wave.
We first introduce the coordinates: $X^M=(t,z,y^m,\tilde{t},\tilde{z},\tilde{y}_m)$  so the wave will be oriented in the $t,z,\tilde{t},\tilde{z}$ directions with the $y$ and $\tilde{y}$ directions transverse.
Then the  solution{\footnote{We give DFT solutions as line elements so as to conveniently encode the generalised metric even though the concept of a line element is problematic since it is not $O(d,d)$-invariant.}} maybe written as follows \cite{Berkeley:2014nza}:
\begin{subequations}
\begin{equation}
\begin{aligned}
{\rm d}s^2 &= M_{IJ} {\rm d}X^I {\rm d}X^J \\
&= (H-2)({\rm d}t^2-{\rm d}z^2)-H({\rm d} \tilde{t}^2- {\rm d} \tilde{z}^2)\,+ \\
&\kern.5cm+ 2(H-1)({\rm d}t {\rm d}\tilde{z} +{\rm d} \tilde{t} {\rm d}z)\,+  \\
&\kern.5cm+ {\rm d}y^2 + {\rm d}\tilde{y}^2  \, 
\end{aligned}
\end{equation}
with 
\begin{align}
e^{2d}= {\rm constant} \, .
\end{align}
\end{subequations}
This DFT wave solution may now be examined with an alternative choice of section or equivalently just rotate the direction of propagation to $x$ space. When one does so one recovers the usual wave solution of supergravity. And thus the T-dual objects of wave and string are a single wave solution in DFT with a single charge the DFT momentum. The orientation of the momentum is what determines the interpretation in space-time as a string or wave. The well versed reader will be aware that in general relativity the notion of conserved currents such as energy and momentum require some care to construct and in fact for a generic solution they do not exist. As fitting with Noether's intuition there needs to be a global symmetry for a conserved charge which in turn means a Killing symmetry (at least asymptotically). This allows the definition of the ADM mass and the Komar like integrals. The equivalent to these have been constructed in DFT in \cite{Blair:2015eba,Park:2015bza,Naseer:2015fba} so that when we refer to the momentum of the solution in DFT we are using the prescription given in these papers.

Note, the wave solution is actually not singular even though the string solution is. We thus have a realisation of one of the goals of string theory, to remove singularities from supergravity solutions.

Now that we have done the electric charges, the next step is to find the magnetic charges. We will follow the Gross--Perry--Sorokin ansatz and fibre a circle around a transverse $S^2$ as before to form the Hopf fibration of $S^3$. We will now take the fibre to be in the $\tilde{x}$ space and use the same coordinate system as before:
\begin{subequations}
\begin{equation}
\begin{aligned}
{\rm d}s^2&= M_{MN} {\rm d}X^M {\rm d}X^N  \\
&= H(1+H^{-2}A^2){\rm d}z^2 + H^{-1} {\rm d} \tilde{z}^2\,+  \\
&\kern.5cm+ 2H^{-1} A_i({\rm d}y^i {\rm d}\tilde{z}- \delta^{ij} {\rm d}\tilde{y}_j {\rm d}z)\,+ \\
&\kern.5cm+ H(\delta_{ij} +H^{-2} A_i A_j){\rm d}y^i {\rm d}y^j +H^{-1} \delta^{ij} {\rm d} \tilde{y}_i {\rm d} \tilde{y}_j\,+ \\
&\kern.5cm+ \eta_{ab} {\rm d}x^a {\rm d}x^b + \eta^{ab} {\rm d}\tilde{x}_a {\rm d} \tilde{x}_b \, .
\end{aligned}
\end{equation}
The DFT dilaton is now:
\begin{align}
e^{-2d}= H e^{-2 \phi_0} \, ,
\end{align}
\end{subequations}
and $A_i$ and $H$ are related as with the Gross--Perry--Sorokin monopole. For more details see the paper \cite{Berman:2014jsa}.
When we examine this solution with the usual choice of section i.e. $x$ space then it is the NS five-brane solution with magnetic charge:
\begin{align}
Q_M= \int H
\end{align}
which is the integral of the $H$-flux over a transverse three cycle ($H$ should not be confused with the harmonic function appearing in the metric).
This is no surprise as the five-brane is the magnetic dual of the string. The solution with the alternative choice of section produces the Kaluza--Klein-monopole. The monopole and five-brane are T-dual so again we are producing a single solution where different choices of section give T-duals. 

Now the attentive reader may worry we have come perilously close to geometrising the gerbe. In the sense of producing a total geometric space for non-trivial gerbe. This is in general not possible. This construction does produce a three-form flux in cohomology but does so by also requiring an additional circle so that one can use:
\begin{align}
H^3(S^1 \times S^2)= H^2(S^2) H^1(S^1) \, ,
\end{align}
and the usual monopole construction to form $H^2(S^2)$ through a Hopf fibration. Thus in fact one is not producing the general NS five-brane but a smeared NS five-brane on a circle. Making more general fluxes and also `localising' the brane on the circle is something that leads to so called `winding mode corrections' of the solution.  Such corrections were first predicted back in 1998 by Gregory, Harvey and Moore \cite{Gregory:1997te} and then realised in DFT in a series of works \cite{Jensen:2011jna,Kimura:2013zva,Berman:2014jsa,Lust:2017jox,Kimura:2018ain}. This is one of the applications of DFT, one simply allows the delta function source of the harmonic function to be localised in winding space and one recovers a harmonic function that was winding mode dependencies. Those are the string theory world-sheet instanton corrections.
This works because windings of string are dual to momentum modes and so in the right variables (those of DFT) they may be modeled with a Poisson equation.

Finally, in order for this to work we require the $\tilde{x}$ direction to be a circle and then the momentum in this direction must be quantised and so we then get the usual Dirac quantisation for $p$-branes.
With the $\tilde{x}$ direction with radius $\tilde{R}$ and the momentum quantised in units of 
$\frac{1}{\tilde{R}}$ this then gives the string tension as $\frac{1}{\alpha'}= \tilde{R}^{-2}$. Remarkably then the string tension is spontaneously generated from compactification i.e.~having $\tilde{x}$ with a scale. One might ask what happens in the decompactification limit where $\tilde{R} \rightarrow 0$. In this case the string tension would go to zero and one is tempted to think we are describing string theory in a tensionless phase. Equally if $\tilde{R} \rightarrow \infty$ then the string tension goes to infinity and we only have supergravity.

But what more? Now we have an extended theory all be it with a constraint. We can non-trivially fibre the new novel directions while still obeying the constraint to make non-trivial fluxes.
Sometime ago people \cite{Shelton:2006fd} looked at what happens when one T-dualises $H$-flux and so called geometric flux (the measure of the geometrical twist). This was done with a toy model of the three-torus with $H$-flux and the so called twisted torus. They are toy models because they are not solutions of supergravity (or string theory low energy effective) equations of motion. This gave rise to the following sequence of fluxes:
\begin{align}
H_{abc} \rightarrow f^a{}_{bc} \rightarrow Q^{ab}{}_c \rightarrow R^{abc}~,
\end{align}
where $H$ and $f$ are the usual three-form and geometric fluxes in supergravity. $Q$ flux seemed something new and $R$ flux even more exotic and contentious since it seemed to require a T-dualisation in a non-isometric direction. Looking at actual solutions in supergravity, the $H$ flux was sourced by an NS5-brane and the $f$-flux by the Kaluza--Klein monopole. The $Q$-flux was shown to be sourced by an exotic brane \cite{deBoer:2012ma} known as the $5^2_2$ brane which although it locally  solves supergravity equations of motion it requires a patching with an $O(d,d)$-transformation \cite{Hassler:2013wsa}.  As such, exotic branes although solutions of string theory are not globally defined solutions of supergravity. They are of course bona-fide solutions of DFT. Branes source R-flux can also be found but they are even more exotic as they require a non-canonical choice of section condition so that there is $\tilde{x}$ dependence in the solution. See \cite{Plauschinn:2018wbo} for an excellent review of non-geometric fluxes.

In the case of usual Kaluza--Klein theory, one could consider more sophisticated reductions such as the\linebreak Scherk--Schwarz-type where one introduces twist matrices to allow non-Abelian gauge fields in the reduced theory. One can do this also with DFT and have generalised Scherk--Schwarz reductions. This leads to gauged supergravity where the gauging is related to the twist field. This is described in a series of works \cite{Aldazabal:2011nj,Geissbuhler:2011mx,Grana:2012rr,Berman:2013cli}. These generalised Scherk--Schwarz reductions are one of the key applications of double field theory. They allow gauged supergravities that otherwise would have no lift or origin from a higher-dimensional theory.

\section{Lifting $11d$ supergravity,  exceptional field theory}

Kaluza--Klein theory incorporates vector fields into the metric in one dimension higher. Double field theory incorporates the two-form $B$-field into a generalised metric of a space with twice the dimension. What about other $p$-form potentials such as the RR forms of type theories or the three-form and six form  of eleven-dimensional supergravity? The RR sector of string theory lifts to eleven-dimensional supergravity as described in our description of M-theory above so if we can lift the $p$-form potentials of eleven-dimensional supergravity then we can do all of string theory. We can also come at this question from an even more formal direction. Usual gravity can be though of as the description of the $GL(d)/SO(d)$ coset. This is most conveniently seen in the vierbein formalism where the vierbein has one $GL(d)$ curved space index and one $SO(d)$ tangent space index. Generalised geometry is concerned with the $O(d,d)/O(d) \times O(d)$ coset. The generalised metric of the previous section parameterises this coset. What then for other cosets? Recently \cite{Morand:2017fnv} has explored having $O(d,d)$ cosets but with different possibly asymmetric tangent space groups. This leads to Double field theory describing a non-Riemannian space. Now, can we ask about cosets of the exceptional groups? Remarkably, exceptional cosets will provide us with the right geometry for eleven-dimensional supergravity. In fact one can then follow \cite{Morand:2017fnv} and investigate different exceptional cosets leading to a host of non-Riemannian geometries in M-theory \cite{Berman:2019izh}. One unfortunate property of the exceptional groups is that typically one has to deal with them on a case by case basis rather than being able to make a general statement for $E_d$.

To make progress, we will now consider a simple example that will illustrate most of the issues involved and study the lifting of a three-form potential in four dimensions to a higher-dimensional space. (Given that we started with a one-form of Kaluza--Klein and then the two-form of NS-NS supergravity, the three-form theory is the obvious next step, the restriction to $4d$ makes the embedding in M-theory natural as will be discussed later).

Lets follow the same process as before. The field content is the metric $g_{\mu \nu}$ and the three-form which we denote by $C_{\mu \nu \rho}$ and since we are in four dimensions $\mu, \nu=1,\ldots,4$. The gauge transformation of $C$ is the usual one for an Abelian p-form, $\delta C_{(3)}= {\rm d} \chi_{(2)}$, and its field strength given by the exterior derivative of $C$ is called $H={\rm d}C$. So as before we want to combine this local symmetry with diffeomorphism onto one `generalised diffeomorphism'. The generalised diffeomorphism will be generated by a generalised vector $V^I=(v^\mu, \chi_{[\mu \nu]})$. Where $I=1,\ldots,10$, since there are four component of $v^\mu$ but six components of $\chi_{[\mu \nu]}$.  One now makes the move following what we did in double field theory and introduce coordinates $\{ x^\mu, y_{[\mu \nu]} \}$ that follows the same structure. 

We combine these coordinates to produce an $SL(5)$ representation. The usual coordinates $x^\mu$ and new novel $y_{[\alpha \beta]}$ coordinates become the {\bf{10}} of $SL(5)$ which we denote with the indices $I,J=1,\ldots,10$. Associated to these coordinates will be a set of translation generators or generalised momenta: $P_I $. 

One then writes down a metric on this ten dimensional space in terms of the usual metric $g_{\mu \nu}$ and three-form $C_{\mu \nu \rho}$ as follows:
\begin{subequations}
\beq
G_{IJ}= (\det(g))^{-1/2}\left( \begin{array}{ll} g_{\mu \nu}+\frac{1}{2}C_\mu{}^{\epsilon \tau}C_{\nu \epsilon \tau}&
\frac{1}{\sqrt{2}}C_\mu{}^{\sigma \rho}\\ \frac{1}{\sqrt{2}}C_\nu{}^{\gamma \delta} & g^{\gamma \delta,\sigma \rho}
\end{array} \right)~,
\eeq
where
\beq
g^{\mu \nu,\sigma \delta}=\tfrac{1}{2} (g^{\mu \sigma}g^{\nu \delta}-g^{\mu \delta}g^{\nu \sigma}) \, ,
\eeq
\end{subequations}
is the induced metric on two-forms. In fact this generalised metric is a representation of the coset:
\beq
SL(5)/SO(5)  \, .  
\eeq
The reader worried about the overall factor of $\det(g)$ in the generalised metric is encouraged to read \cite{Berman:2011jh} and \cite{Malek:2012pw} where the important role of this factor in U-duality is discussed. 

Note, that as with DFT this is now a metric not on the usual tangent bundle as with Riemannian geometry but on:
\beq
TM \oplus \Lambda^2 T^*M  \, .
\eeq

Again one can see a Kaluza--Klein style of structure with the $C$-field being the off diagonal components mixing the usual space with the new novel space.

The next step is to find an action that reduces to the one for gravity and a three-form once one removes the dependence of the novel coordinates $y_{[\mu \nu]} $ from the fields. That is when $\partial^{[\mu\nu]} \cdot =0$ the action will becomes Einstein--Hilbert with an appropriate $H^2$ term.
The action that does this is \cite{Berman:2010is}:
\begin{equation}
\begin{aligned}
S&=   \tfrac{1}{12} M^{MN} \partial_M M^{PQ} \partial_N M_{PQ} - \tfrac{1}{2} M^{MN} \partial_M M^{PQ} \partial_P M_{NQ}\,+ \\ 
&\kern.5cm+ \tfrac{1}{84} M^{MN}(M^{KL} \partial_M M_{KL} )(H^{PQ} \partial_N H_{PQ} )    \, .
\end{aligned}
\end{equation}

So having constructed the coordinates, the generalised metric on the space and the action, the next step is to look at the local symmetries. The local symmetries are given by the generalised Lie derivative  as in equation (\ref{genlie}). Now the Y-tensor must be constructed from $SL(5)$ invariant tensors and under the constraint $\partial^{[\mu\nu]} \cdot =0$ the local transformations must become diffeomorphisms and three-form gauge transformations. This is achieved with the Y-tensor being given by: $Y^{IJ}{}_{PQ}=  \epsilon^{IJ i}\epsilon_{PQ i}$ with $I,J$ indices being in the {\bf{10}} of $SL(5)$ and $i$ being a {\bf{5}} of $SL(5)$. Given the generalised Lie derivative one can now seek closure of the algebra. Just as in the DFT case, the algebra will not close unless:
\beq
Y^{IJ}{}_{PQ} \partial_I  \partial_J  =0  \, .
\eeq
A full discussion of the Exceptional Field Theory (EFT) local algebra is given in \cite{Berman:2012vc}.
One can then show with some work that there are two independent solutions to this condition \cite{Blair:2013gqa} up to trivial transformations. One is where we remove the dependence on the novel coordinates leaving us with the original $x^\mu$. The other is where we keep dependence on $y_{14},y_{24},y_{34}$. These choices are related to eleven-dimensional supergravity and IIB supergravity respectively.
This says that now when we make different choices of section (by this we mean choosing the coordinates that we identify as being space-time) we can obtain either $11d$ or IIB in $10d$. This is not a surprise since exchanging winding and moment in M-theory does not preserve dimension (essentially because the membrane has two spatial dimensions). Different choices of section in DFT in the presence of isometries led to T-duality. Now different choices of section lead to the U-duality transformations; again one must have sufficient number of isometries to realise the U-duality group. 

The astute reader will be at this point trying to work out where the eleven (or indeed ten dimensions) are since we have only managed to get a theory in four (or three) dimensions. The way one thinks about this is as follows. Take $M^{11}=M^4 \times M^7$. At this point the $M^7$ will be entirely trivial e.g.~a seven torus which nothing depends on. Then we augment the four-dimensional space with the additional 6 novel dimensions to give 10 dimensions described by $SL(5)$ generalised geometry and 7 dimensions described by usual Riemannian geometry. Therefore we have 17 dimensions in total before then making a section choice to go down to eleven or ten dimensions. This is the general story. 
We split the eleven-dimensional space as $M^{11}=M^d \times M^{11-d}$ and then add novel coordinates to $M^d$ to make them a representation of $E_d$. The generalised Lie derivative will be the same, only the Y-tensor will change (this is true up to and including $E_7$).  The section condition coming from the closure of the algebra will also be the same once written in terms of the Y-tensor (again up to and including $E_7$). 

The next steps then are to look at the states of the theory just as we did in Kaluza--Klein theory and DFT. Null momenta in the novel $y_{\mu \nu}$ directions give a membrane solution wrapped on the $\mu \nu$ directions. Thus we associate the new novel directions with membrane winding modes.

The next step is to consider the new coordinates fibred non-trivially over the space-time base. A Kaluza--Klein type monopole solution whose Hopf fibre has coordinates $y_{[\mu \nu]}$ reproduces a wrapped five brane. This is exactly keeping with the intuition from previous cases where the fibred solution gives the magnetic state and the state with momentum gives the electrically charged state.

Finally, the reader will have noticed we have rather brutally truncated the theory by completely ignoring the $M^{11-d}$ space. What is now called exceptional field theory as developed by Hohm and Samtleben \cite{Hohm:2013pua,Hohm:2013vpa,Hohm:2013uia,Hohm:2014fxa} puts this back in and allows for arbitrary fibrations of one space over the other. The space we have been dealing with goes by the name `internal space' and the here neglected Riemannian space is called the `external space'. The are field that have indices both spaces most importantly a field $A^I_a$ which is a one-form in the `external space' and a vector in the `internal space' and describes how one is fibred over the other. The local symmetries need to be made to be consistent between both spaces. All these considerations lead to a highly complex theory that is described in a series of works initially by Hohm and Samtleben \cite{Hohm:2013pua,Hohm:2013vpa,Hohm:2013uia,Hohm:2014fxa} and then by others \cite{Hohm:2015xna,Abzalov:2015ega,Musaev:2015ces,Berman:2015rcc}.

The relevant coset, $G/H$ for a given dimension $d$ of the internal space are listed in the table \ref{GHR} along with $R_1$ which is the coordinate representation of $G$. We have also listed $H^*$ for when the internal space is Lorentzian. 

\begin{table}\centering
\begin{tabular}{ccccc} 
$d$ & $G$ & $H$ & $H^*$  & $R_1$  \\\hline 
4 & $\Gfour$ & $\Hfour$ & $\HfourLor$  & $\mathbf{10}$  \\
5 & $\Gfive$ & $\Hfive$ & $\HfiveLor$ & $\mathbf{16}$ \\
6 & $\Gsix$ & $\Hsix$ & $\HsixLor$ &  $\mathbf{27}$ \\
7 & $\Gseven$ & $\Hseven$ & $\HsevenLor$& $\mathbf{56}$ \\
8 & $\Geight$ & $\Height$ & $\HeightLor$ & $\mathbf{248}$ \\
\end{tabular}
\caption{The cosets $G/H$ and the dimension of the coordinate representation for dimension $d$, internal space.} 
\label{GHR} 
\end{table} 

Before moving on lets consider some of the cases in other dimensions and gain some intuition for the table of cosets.
Let us see what happens in the next instance of $d=5$. We will now reverse the logic from the above. In the previous instances we looked at how to lift the local symmetries, form an action and then find solutions that correspond to strings and branes in the space. The novel coordinates were then related to winding modes of these branes. Now we will use our knowledge of branes in M-theory to establish what to expect and then one can follow the same procedure as before. M-theory has membranes and five-branes. When we consider the d dimensional internal space we must ask what branes can winding in this space. Up to 5 dimensions it is only the membrane and so the coordinates $x^{\mu},y_{[\mu \nu]}$ describe this. But when we have 5 dimensions or more then we must include coordinates for wound five-branes thus augmenting the previous set with $y_{[\mu \nu \rho \sigma \tau]}$. There is then one more subtlety, which is that we can have wound D6-branes which from the eleven-dimensional perspective is a Kaluza--Klein monopole with 6 world-volume directions and one Hopf fibre direction and thus it would have 6 antisymmetrised coordinates and a Hopf coordinate. When one takes this into account then one gets the dimension of the coordinate representations in the table.

Once we have sufficient number of dimensions to include the five-brane winding coordinates something new happens when we consider solutions. We have coordinates for both the wrapped membrane and the wrapped five-brane but these are electromagnetic duals of each other. Thus following the previous logic there would be two ways to get the same solution. One could describe the five-brane as a null wave in the $y_{[\mu \nu \rho \sigma \tau]}$ direction or as a monopole with Hopf fibre given by $y_{[\mu \nu]}$. In fact since electromagnetic duality is contained in the U-duality group these must be the same solution in the exceptional field theory. In fact one would then make the same argument for the membrane and describe it as either a null wave in the $y_{[\mu \nu]}$ direction or as a monopole with Hopf fibre given by $y_{[\mu \nu \rho \sigma \tau]}$. The answer is that the solution must be `self-dual' meaning that it will be a wave in one direction and a monopole in the electromagnetic dual direction. This is reported in detail in \cite{Berman:2014hna} where these self-dual solutions to EFT were constructed and shown to describe the branes in M-theory and their bound states. For those familiar with the (2,0) theory in 6 dimensions where a self-dual string solution gives rise to the states in 4 dimensions that transform under the 4d $SL(2)$ after dimensional reduction on a torus, this is analogous. The solution described in \cite{Berman:2014hna} is a sort of gravitational version of the self-dual string. Its reduction in different ways gives rise to the brane states that transform under the U-duality group. One might suspect that this is always true. If we have a theory with states transforming under some duality group it suggests that the theory can be lifted to higher dimensions and that these states come from a single solution in the higher-dimensional theory.

Following the Kaluza--Klein intuition leads to the idea of a Scherk--Schwarz-type ansatz for reducing EFT \cite{Berman:2012uy,Musaev:2013rq,Aldazabal:2013mya}. This breaks the usual section condition and yet one can show it is consistent. The result is produces gauged supergravities where the gauging is determined by the so called embedding tensor which in turn is given by the twist matrix of the Scherk--Schwarz ansatz.

Finally, we can consider more complicated fibrations and use the full $E_d$-symmetry to patch solutions. This leads to M-theory generalisations of the exotic fluxes described for DFT and the appropriate exotic brane solutions that act as their source. This is reported in depth in \cite{Bakhmatov:2017les, Fernandez-Melgarejo:2018yxq, Berman:2018okd} where a huge spectrum of solutions have been constructed using EFT.

\section{Superalgebras}
(The work in this section is based on a collaboration with Malcolm Perry \cite{BermanPerry}).
Perhaps the starting point for M-theory is the fact that the type II superalgebras in ten dimensions can be lifted to the unique eleven-dimensional superalgebra.

\beq
\begin{aligned}
&\{ Q_\alpha, Q_\beta \}=\\
&\kern.5cm= P_\mu (C\Gamma^\mu)_{\alpha \beta} + Z^{\mu \nu} (C \Gamma_{\mu \nu})_{\alpha \beta} + Z^{\mu_1\cdots\mu_5}(C \Gamma_{\mu_1\cdots\mu_5})_{\alpha \beta}  \label{11dsusyalg}
\end{aligned}
\eeq 
In lifting the type IIA algebra we must identify the ten dimensional central charge associated to the D0-brane with the momentum in the eleventh dimension. 
This identification can be seen directly by comparing the  BPS state equation:
\beq
P_0{}^2=|Z|^2  \label{BPS}
\eeq
to the equation for a null wave:
\beq
P_0{}^2=|\vec{P}|^2  \, .
\eeq
The D0 brane identification with the  null wave in M-theory exactly realises this formal similarity with the central charge, $Z$ being identified with $P_{11}$.
Thus the D0-brane is actually massless. Its effective mass in ten dimensions is just a result of its momentum in the eleventh dimension. Looking at the eleven-dimensional superalgebra (\ref{11dsusyalg}) it is tempting to see if one could try the same trick again. Can one could reinterpret all the central charges as arising from momenta in extra dimensions? If so then all the branes would result from null-waves in extra dimensions i.e.~all the branes would be tensionless and their effective tensions in eleven dimensions would only arise from their momenta in the extra dimensions. This is not a new idea. A  number of authors have pursued this idea in various forms. In fact the seed of this idea was noticed at the very naissance of extended supersymmetry in where the central charges of the $\mathcal{N}=2$,  $4d$ theory were described as coming from some higher-dimensional theory. 

The reader tempted by this idea will immediately be put off by the fact that the central charges (\ref{11dsusyalg}) transform as a Lorentz two-form and five form. Also the objects contracted with the central charges i.e. $C\Gamma_{\mu \nu}, C\Gamma_{\mu_1\cdots\mu_5}$ do not appear to obey a Clifford algebra!
This immediately seems to end the idea of interpreting the central charges as momenta and alternative interpretations were explored for the so called {\it{M-theory algebra}}  \cite{Hull:1994ys,Sezgin:1996cj}.

However, this is exactly what the sort of extended generalised geometry described above is set up to do where the {\it{generalised}} coordinates in the extended geometry are (we will only consider $d<6$):
\beq
X^I=(x^\mu, y_{\mu  \nu}, y_{\mu_1\cdots\mu_5} ) \, . \label{extracoords}
\eeq

Given the discussion above it seems natural to revisit the idea of lifting the eleven-dimensional superalgebra and interpreting the central charges as momenta in these new dimensions. We will thus examine some specific examples with dimension, $d=10,16$  with a simple superalgebra without central charges, that is schematically:
\beq
\{Q_\alpha, Q_\beta \} = (C\Gamma_I)_{\alpha \beta} P^I  \, .  \label{susynocentral}
\eeq 
We will then examine the representation theory for this superalgebra. The {\it{massless}} (in the generalised sense) representations will then be shown to obey quadratic constraints on the momenta in the theory,
\beq
Y^{IJ}{}_{PQ} P_I P_J =0 \,  , \quad P^IP^J  \delta_{IJ}=0 \, .
\eeq
 The first quadratic constraint is exactly the {\it{physical section condition}} that was discovered previously from demanding closure of the algebra a local symmetries  \cite{Berman:2011cg,Coimbra:2011ky,Berman:2012vc} and the second condition is just the statement of being massless (in the generalised theory). Other representations will not obey this constraint and so will not be contained in the usual formulation of generalised or extended geometry where the section condition must be imposed. 
In terms of the usual theory these generalised massless states will be the 1/2 BPS states. The section condition determines a generalised {\it{light cone}} where the 1/2 BPS lie, the other states with less supersymmetry then lie in the interior of the cone. 

This shows the remarkable interplay between the global duality symmetry, the local symmetries and the supersymmetry in the extended space.
This quadratic constraint on momenta restricts the effective degrees of freedom and avoids the various no-go theorem for constructing supersymmetric theories in dimensions higher than eleven.

\subsection{U-duality and its realisation in an extended geometry, the simple example of $SL(5)$}

Along side the usual four coordinates, $x^a$, ($ a =1,\ldots,4$), six new coordinates $y_{[ab]}$ are introduced to make up a ten dimensional space.  (We move here to Latin indices to indicate we must consider different possible signatures).
The key to making supersymmetry work is that the spinors live in the local group H. Thus for the $SL(5)$ U-duality invariant action the bosonic sector is given by a nonlinear realisation of $SL(5) / SO(5)$ but the spinors of the theory will be a representation of ${\rm Spin}(5)$.

We now recombine the coordinates to produce an $SL(5)$ representation. The usual coordinates $x^a$ and the membranes windings $y_{ab}$ become the {\bf{10}} of $SL(5)$ as follows:
\beq
\begin{aligned}
z^{[ij]}&= z^{a5 } = x^a \\  
            &= z^{ab} = \tfrac{1}{2} \eta^{abcd} y_{cd}  \, \, . \label{sl5coords}
            \end{aligned}
\eeq
Where $\eta^{abcd}$ is the alternating symbol, i.e. $\eta^{1234}=1$ and
$i,j=1,\ldots,5$ and the coordinates $z^{[ij]}$ are in the {\bf{10}} of $SL(5)$ which we denote with the index $I,J=1,\ldots,10$. Associated with these coordinates will be a set of translation generators or generalised momenta: \beq P_{[ij]}=P_I  \, .
\eeq 

One then writes down a generalised metric on this ten dimensional space in terms of the usual metric $g_{ab}$ and three-form $C_{abc}$ as follows:
\begin{subequations}
\beq
G_{IJ}= \Bigl( \begin{array}{ll} g_{ab}+\frac{1}{2}C_a{}^{ef}C_{bef}&
\frac{1}{\sqrt{2}}V_a{}_{kl}\\ \frac{1}{\sqrt{2}}V_b{}_{mn} & g_{mn,kl}
\end{array} \Bigr)~,
\eeq
where
\beq
g_{mn,kl}=\tfrac{1}{2} (g_{mk}g_{nl}-g_{ml}g_{nk})
\eeq
is the induced metric on antisymmetric bi-vectors and we have defined $V_a{}_{kl}$ to be given by
\beq
V_a{}_{kl}= C_a{}^{pq} \eta_{klpq}  \, .
\eeq
\end{subequations}
The  metric $G_{IJ}$ is a metric on the coset of $SL(5)/SO(5)$. This is just a more convenient rewriting from the previous description of the $SL(5)$ theory that is suited to what follows.

In order to make the theory supersymmetric we will require that the space have a Lorentzian signature and thus $g_{ab}$ is Lorentzian. The reason for this, aside from any desire to construct a theory with a temporal direction, simply comes from the usual restriction of spinors in various dimensions and signatures \cite{Freedman:2012zz}. Thus instead of the usual Euclidean 14-dimensional coset with (0,14) signature we will need to work with the coset:
\beq
SL(5)/SO(2,3) \, .        \label{sl5coset}
\eeq
This coset's dimension\footnote{Not to be confused with the dimension of the space itself on which this metric acts which is of course 10.} is 14 with signature given by (10,14)-(4,6)=(6,8). The various choices of Lorentzian\linebreak coset structures are discussed in \cite{Malek:2013sp}.
To see why this is the appropriate choice of coset we simply examine the fields in the coset and count the number with negative directions. These negative directions are given by: $g_{0 a}$ and $C_{0ab}$ (with $a=1,\ldots,3$) which gives a total of six negative directions leaving 8 positive directions which indeed matches the counting of the coset given in (\ref{sl5coset}).

We will now introduce spinors of the local group $H$, $SO(2,3)$,  with spinor index $\alpha=1,\ldots,4$. Consequently we then have the associated gamma matrices:
 \beq (\Gamma^i){}^{\alpha}{}_ {\beta} \ , \qquad  i=1,\ldots,5 \eeq which form the Clifford algebra for $SO(2,3)$. Along with this we have the charge conjugation matrix $(C)_{\alpha \beta}$ and its inverse $(C^{-1})^{\alpha \beta}$ with which we can lower and raise spinor indices respectively through left multiplication.

From these $SO(2,3)$ $\Gamma$ matrices we can form the appropriate representation of the global group G, $SL(5)$. The set of antisymmetrised products of the $\Gamma^i$ matrices:
\beq
(\Gamma^{[ij]})^\alpha{}_\beta = (\Gamma^I)^\alpha{}_\beta  \, , \qquad I=1,\ldots,10
\eeq
 are in the {\bf{10}} of $SL(5)$. To compare with the usual supersymmetry algebra, they can be decomposed into $SO(1,3)$ $\Gamma$ matrices just as we did with the coordinates described by (\ref{sl5coords}):
\beq
\Gamma^I=\Gamma^{[ij]}=(\Gamma^{[a5]},\Gamma^{[cd]})  \, . \label{sl5gammas}
\eeq
Similarly, the generalised momentum decomposes as:
\beq
P_I=P_{[ij]}=(P_{a5}, \tfrac{1}{2} \eta_{abcd} Z^{cd})
\eeq
with the obvious identification of $P_{a5}=P_a$ with momenta in the usual four-dimensional space-time and the set $\{Z^{cd}\}$ labels momenta in the novel extended directions.
Now we wish to be able to form a supersymmetry algebra using this set of generalised gamma matrices, $\{ \Gamma^I \}$, the set of generalised momenta $P_I$ and the supercharges, $Q_\alpha$. No central charges are required; the bosonic sector has only the generators of the generalised Poincare group.
Thus the complete superalgebra is given by:
\beq
\{Q_\alpha,Q_\beta\}= (C\Gamma^I)_{\alpha \beta} P_I \, ,     \qquad [Q_\alpha,P_I]=0   \, , \qquad [P_I,P_J]=0 \, ,
\eeq
where $C$ is charge conjugation matrix for $SO(2,3)$ spinors.
The is supplemented by the Lorentz algebra for $SO(2,3)$ which will act not on the vector representation but on the $\bf{10}$ of $SO(2,3)$ so that when combined with the momenta $P_I$ we have the generalised Poincare group i.e. the motion group for the generalised space-time. Just as in any quantum field theory we will use the algebra of space-time to classify states. Following Wigner, elementary states are irreducible representations of the Poincare algebra and we may use the Casimirs of the algebra to label the representation. In this case it is the generalised Poincare algebra that will be relevant to classify the states of the theory through its Casimirs.

We will proceed exactly as in the usual superalgebra case when one wishes to examine the massless representations i.e. where the quadratic Casimir of momentum vanishes, and show that they form`short multiplets'.

We calculate the square of (\ref{susynocentral}) which is positive definite:
\beq
(C^{-1}\Gamma^I )^{\alpha \beta}\, (C \Gamma^J)_{\beta \gamma} P_I P_J  \geq 0   \, .  \label{susyaxiom}
\eeq

Then by demanding that this bound is saturated, we have a quadratic constraint on the generalised momenta, $P^I$.

We will now determine this constraint on the generalised momenta by substituting the decomposition (\ref{sl5gammas}) into (\ref{susyaxiom}) and demanding the bound is saturated.

This produces (suppressing spinor indices):
\beq
\begin{aligned}
&( C^{-1}\Gamma^I )\, (C \Gamma^J) P_I P_J =\\
&\kern.5cm= 2\big( \eta_{ab} P^a P^b-Z_{ab} Z_{cd} (\eta^{ac}\eta^{bd}-\eta^{ad}\eta^{bc})   \big) \mathbbm{I}\,+  \\
&\kern1cm+ 4 P^a Z_{ab} \Gamma^b+  \\   
&\kern1cm+ 2 Z_{ab} Z_{cd}   \Gamma^{abcd}  \, . 
\end{aligned}
\eeq

Demanding that this is zero, we need each line to vanish separately. This means the constraints in terms of the four-dimensional momenta and central charges are:
\beq
P_aP^a = Z_{ab} Z^{ab} \, , \qquad P^a Z_{ab}= 0   \, ,  \qquad \,  Z_{ab} Z_{cd} \epsilon^{abcd} =0 \, .
\eeq
The first term is the standard BPS condition (\ref{BPS}) requiring the mass be equal to the central charge and the second two equations are the quadratic constraints required for the state to be 1/2 BPS as calculated in \cite{Obers:1998fb} by essentially the same calculation.

In terms of the $SL(5)$ generalised momenta, $P_{[ij]}$ these equations become:
\begin{subequations}
\bea
&P_{[ij]} P^{[ij]}= P_I P^I=0 \,  , \\  \label{bpssl5}
&\epsilon^{ijklm} P_{[ij]} P_{[kl]} =0  \,  . \label{scsl5}
\eea
\end{subequations}

The first equation (\ref{bpssl5}) implies that the state is massless in from the point of view of the extended space Poincare algebra. Thus in extended geometry the usual BPS states are massless. Supersymmetry works because the massless multiplet of $SO(2,3)$ has the same number of degrees of freedom as the massive multiplet in $SO(1,4)$.
\medskip

The second equation (\ref{scsl5}) is precisely the {\it{physical section condition}} that we need to impose so that the local symmetry algebra of the extended geometry i.e. the algebra of generalised Lie derivatives, closes. 

Thus we see that from demanding the representation of the supersymmetry algebra saturates the bound (\ref{susyaxiom}) we reproduce the quadratic constraints on the generalised momenta. The foundation of this calculation has essentially already appeared in the literature in the context of U-duality multiplets for 1/2 BPS states \cite{Obers:1998fb}\footnote{We thank Boris Pioline in particular for first drawing our attention to the similarity between the physical section condition and the 1/2 BPS constraints.} and more recently the condition has been rewritten in terms of the $E_{11}$ algebra in \cite{West:2012qm}. What this calculation shows is the connection between: 1/2 BPS states in 4 dimensions, these states in the 10-dimensional extended space, the spinors of $SO(2,3)$, the local Lorentz group of the extended space and their Clifford algebra, and representations of $SL(5)$ the global symmetry of the extended space. 

Finally, we now rewrite the supersymmetry algebra of the generalised space with no central charges  (\ref{susynocentral})  in terms of four-dimensional quantities i.e. 4d momenta and central charges as follows:
\beq
\begin{aligned}
\{Q_\alpha, Q_\beta \} &= (C\Gamma^II)_{\alpha \beta} P_I \\
&= (C  \Gamma^a \Gamma^5)_{\alpha \beta} P_a + \big(\tfrac{1}{2} \eta_{abcd}  C \Gamma^{ab}\big)_{\alpha \beta}   Z^{cd}  \, . \label{5to4}
\end{aligned}
\eeq
Now we wish to think of this in $SO(1,3)$ language so that we can reinterpret the spinors as being Dirac spinors of $SO(1,3)$. Obviously, the Spin$(2,3)$ spinors also can be thought of as Spin$(1,3)$ Dirac. Crucially, the charge conjugation matrix will be different because of the presence of two time like directions in $SO(2,3)$. Thus the four dimension $SO(1,3)$ charge conjugation matrix which we denote by $C_{(4)}$ will be related to the $SO(2,3)$ charge conjugation matrix by:
\beq
C=C_{(4)} \Gamma^5
\eeq
with $(\Gamma^5)^2= -1$.
We can then insert this into (\ref{5to4}) to give:
\beq
\begin{aligned}
\{Q_\alpha, Q_\beta \} &= (C\Gamma_I)_{\alpha \beta} P^I\\
&= (C_{(4)} \Gamma^5 \Gamma^a \Gamma^5)_{\alpha \beta} P_a+ \big(\tfrac{1}{2} \eta_{abcd}  C_{(4)} \Gamma^5 \Gamma^{ab}\big)_{\alpha \beta}   Z^{cd}  \\
&= (C_{(4)} \Gamma^a )_{\alpha \beta} P_a + (  C_{(4)}  \Gamma_{cd})_{\alpha \beta}   Z^{cd} \, ,
\end{aligned}
\eeq
where we have used the elementary properties of four-dimensional $\Gamma$ matrices:
\beq
\{\Gamma^5,\Gamma^a \}=0 \, \, { \quad \rm{and}} \quad \Gamma^5 \Gamma^{ab} =- \tfrac{1}{2}\eta^{abcd} \Gamma_{cd} \, .
\eeq

Thus, we have seen how the usual $4d$ supersymmetry algebra with central charges may be lifted to an algebra with global $SL(5)$-symmetry with no central charges. All the central charges come from momenta in the novel extended directions. The section condition of the extended geometry is then just from considering the massless representations of this algebra --- the massless representation being 1/2 BPS as usual.

This is the structure that we will replicate for the different cosets. In summary, for dimension $d$ there is a coset $G/H$; one then does the following:
\smallskip
\begin{enumerate}[i)]

\item Construct spinors of H and write down the associated Clifford algebra, $\Gamma^i$.

\item  Form a representation of G using a sum of antisymmetrised products of $\Gamma^i$ to give $\Gamma^I$.

\item Combine the momenta and central charges to form a representation of G which we think of as the generalised momenta $P_I$.

\item Rewrite the superalgebra in terms of only the generalised momenta $P_I$ and the set of generalised gamma matrices $\Gamma^I$.

\item  Demand (\ref{susyaxiom}) is saturated for the {\it{massless}} representation to give constraints on $P_I$.

\item This constraint should be the same as that required by the closure of the algebra of generalised Lie derivatives, also known as the section condition.

\end{enumerate}
\smallskip

In case the reader is worried that the $SL(5)/SO(2,3)$ case was somehow degenerate and we got lucky we will now carry out this procedure explicitly for the case of $d=5$, $G=SO(5,5)$ and $H=SO(5,\mathbbm{C})$ and again reproduce the physical section condition for the theory.

\section{Case,  $SO(5,5)/SO(5,\mathbbm{C})$ and manifest U-duality in $d=5$}

We will now follow the above instruction set for the case in $d=5$ of $SO(5,5)/SO(5,\mathbbm{C})$. The algebra $H$ is now $SO(5,\mathbbm{C})$ rather than the more customary $SO(5)\times SO(5)$ because again we will choose a metric with Lorentzian signature and so the local group, $H$ is different from the Euclidean case. One discovers that the coset has dimension 25 with signature (15,10) as it should. The extended space itself on which the $SO(5,5)$ acts is 16-dimensional with coordinates that we will denoted by $z^I$. In terms of five-dimensional representations:
\beq
z^I=(x^\mu, y_{[ \mu \nu ]}, y_{ [ \mu \nu \rho \sigma \tau ]} ) \, ,~~~~ \mu, \nu=1,\ldots,5
\eeq
that is, usual coordinates augmented with two-form coordinates and five-form coordinates. The metric on this space and the $SO(5,5)$ invariant action are described in \cite{Berman:2011pe}.
 
Step 1,  we construct spinor of $H$. Thus we have complex spinors of $SO(5)$ and the corresponding complex Clifford algebra:
 \beq
(\Gamma^\mu)^\alpha{}_\beta   \, .
\eeq
Step 2, we must now form a representation of $G$ from this set of gamma matrices:
\beq
\Gamma^I=( \Gamma^\mu, \Gamma^{[\mu \nu]}, \Gamma^{[\mu \nu \rho \sigma \tau ]}) \, .
\eeq
Step 3, we write the generalised momenta in terms of a 5d momentum $P^\mu$ and central charges as follows:
\beq
P^I=(P^\mu, Z_{[\mu \nu]}, Z_{[\mu \nu \rho \sigma \tau ]}) \, .
\eeq
Step 4, we demand (\ref{susyaxiom}) to determine constraints on $P^I$, and now we must do some work with the $5d$ Clifford algebra (we have found GAMMA \cite{Gran:2001yh}, very useful for this):
\beq
\begin{aligned}
&\{ C\Gamma^I, C \Gamma^J\} P_I P_J =\\
&\kern.5cm= 2( \eta_{\mu \nu} P^\mu P^\nu+   Z_{\mu \nu} Z^{\mu \nu}  +  Z_{\mu_1\cdots\mu_5} Z^{\mu_1\cdots\mu_5}  ) \mathbbm{I}\,+  \\
&\kern1cm+ 4 P^\mu Z_{\mu \nu} \Gamma^\nu + 2( Z_{[\mu \nu} Z_{\rho \sigma]}   + P^\tau Z_{\tau \mu \nu \rho \sigma}     ) \Gamma^{\mu \nu \rho \sigma} \, .
\end{aligned}
 \eeq
The constraints are thus:
\beq
P^\mu Z_{\mu \nu}= 0 \qquad {\rm{and}} \qquad Z_{[\mu \nu} Z_{\rho \sigma]}  + P^\tau Z_{\tau \mu \nu \rho \sigma} =0  \, .
\eeq
In terms of representations of $G$, this is equivalent to the pure spinor condition on the generalised momenta, $P^I$ and matches precisely the physical section condition \cite{Berman:2015rcc}.

\section{Double field theory and the analogy with geometric quantisation}

We have seen that in DFT we double the coordinates but then later restrict the coordinate dependence on fields in order to solve the strong constraint. Crucially, we do allow alternative solutions of the strong constraint or `section condition' so that the formalism allows us to have fields to be functions of $x$ or $\tilde{x}$ just not both simultaneously. 

This is highly analogous to geometric quantisation \cite{Woodhouse:1992de}, where one has in the pre-quantum bundle a complex line bundle over phase space with $2d$ dimensions. The phase space has canonical  coordinates $\{x,p\}$ and is equipped with a symplectic form $\omega$. Sections of the pre-quantum bundle are wave functions on phase space $\phi(x,p)$.

We must then impose a so called `polarisation' on phase space to take the pre-quantum bundle to the physical quantum bundle which means one determines a Lagrangian submanifold, $\Sigma$ of phase space (that is a d-dimensional isotropic and coisotropic submanifold). The symplectic form is used to determine the polarisation through demanding that the pull back of $\omega$ to $\Sigma$ vanishes. Then one demands that the base of the line bundle is the Lagrangian submanifold. (In more mundane language, that the wave function only depends on the coordinates of the Lagrangian submanifold).

To make all this concrete, let us adopt Darboux coordinates on phase space such that the symplectic form is given by $\omega={\rm d}p \wedge {\rm d}q$. Wave functions $\phi(x,p)$ then can be restricted natural to the Lagrangian submanifold described with coordinates $q$ such that the wave function is independent of $p$ and $\phi(q)$. Note, however, one could make an alternative choice and have the momentum representation such that the Lagrangian submanifold is described by $p$ and $\phi(p)$. These different  choices reflect the hidden symplectic symmetry of the system. Physics is invariant under canonical transformations, which are the coordinate transformations of phase space that preserve the symplectic structure. Note, that in the usual description of the system using a Lagrangian, this symmetry is not manifest. The Lagrangian has the advantage of being naturally relativistic (in treating time and space coordinates on the same footing) but it hides the canonical symmetry that the Hamiltonian exposes.

Now in fact life is not quite as simple as described above. When one makes different choices of polarisation one must (in what is known as `half form'  quantisation) pick a volume form on the Lagrangian submanifold (this is so one can define the norm of the wave function on the space). As one moves between different choices of polarisation then the volume form needs to transform. Further than this, in fact one can allow the wave function to pick up a phase. These changes of phases in the wave function under different choices of polarisation mean in fact that the wave function is not a representation of the symplectic group but its double cover which is known as the metaplectic group \cite{Woodhouse:1992de}.

Note, there is an entirely alternative quantisation scheme, where one keeps the pre-quantum bundle and thus have $\phi(x,p)$ but now deform the product of fields, often called the star product. This is known as Moyal quantisation and has the advantage of maintaining manifest symplectic symmetry under quantisation. There are other quantisation schemes also possible and their equivalence is much discussed.

All of this is highly analogous to DFT and string theory. The phase space of the string contains the coordinates $\{x,\tilde{x}\}$. The strong constraint implies that one must construct a polarisation using $\eta$, i.e. a $d$-dimensional subspace which we call space-time \footnote{In fact one can make a stronger analogy if DFT comes equipped with a complex structure since then it is para-Hermitian and one can extract a symplectic form, \cite{Vaisman:2012ke}.} Alternative choices are possible and the shift between different choices is as we have seen related to T-duality. In fact one can push this analogy further. The shift in the dilaton under T-duality is precisely the sort of shift one sees in half form quantisation of the volume form on the Lagrangian submanifold. Perhaps the most persuasive detail is that the string partition function under T-duality, i.e. changing section choice, transforms as a representation of the metaplectic group. This is easily seen as theta functions transform metaplectically under Poisson resummation. (A fact that was used by Weil to give a representation-theoretic interpretation of theta functions). Essentially from the string world-sheet perspective T-duality is a canonical transformation, the partition function is then like the wave function and DFT is a geometry that reflects all this.

This observation begs the question whether alternative schemes like Moyal quantisation are possible for DFT, where one drops the strong constraint but instead deforms the product of fields to produce string corrections.
In all of this it is implicitly appearant that the string itself is sort of a quantisation, a dimensional deformation of the classical theory equipped with a polarisation on phase space and the structure of  DFT reflects this.

\bibliographystyle{prop2015}
\bibliography{allbibtex}

\end{document}